\documentclass[twocolumn,prl,showpacs,floatfix]{revtex4}

\usepackage{graphicx}
\usepackage{bm}
\usepackage{color}
\usepackage{amsmath}
\usepackage{amsfonts}
\usepackage{amssymb}
\usepackage{epstopdf}

\begin{document}

\preprint{APS/123-QED}

\title{Magnetization Process and Collective Excitations in the $\bm{S=1/2}$ Triangular-Lattice Heisenberg Antiferromagnet Ba$_3$CoSb$_2$O$_9$}

\author{Takuya Susuki$^1$, Nobuyuki Kurita$^1$, Takuya Tanaka$^2$, Hiroyuki Nojiri$^2$,\\
 Akira Matsuo$^3$, Koichi Kindo$^3$, and Hidekazu Tanaka$^1$}
\affiliation{
$^1$Department of Physics, Tokyo Institute of Technology, Meguro-ku, Tokyo 152-8551, Japan\\
$^2$Institute for Material Research, Tohoku University, Aoba-ku, Sendai 980-8577, Japan\\
$^3$Institute for Solid State Physics, University of Tokyo, Kashiwa, Chiba 277-8581, Japan
}
\date{\today}
 
\begin{abstract}
 
We have performed high-field magnetization and ESR measurements on Ba$_3$CoSb$_2$O$_9$ single crystals, which approximates the two-dimensional (2D) $S\,{=}\,1/2$ triangular-lattice Heisenberg antiferromagnet.
For an applied magnetic field $H$ parallel to the $ab$-plane, the entire magnetization curve including the plateau at one-third of the saturation magnetization ($M_\mathrm{s}$) is in excellent agreement with the results of theoretical calculations except a small step anomaly near $(3/5)M_\mathrm{s}$, indicative of a theoretically undiscovered quantum phase transition. However, for $H\,{\parallel}\,c$, the magnetization curve exhibits a cusp near $M_\mathrm{s}/3$ owing to the weak easy-plane anisotropy and the 2D quantum fluctuation.
From a detailed analysis of the collective ESR modes observed in the ordered state,
combined with the magnetization process, we have determined all the magnetic parameters including the interlayer and anisotropic exchange interactions.
  
\end{abstract}

\pacs{75.10.Jm, 75.45.+j, 75.60.Ej, 76.30.-v}

\maketitle


Over the past decades, there has been considerable interest in frustrated quantum magnets,
owing to a rich variety of exotic quantum phenomena\,\cite{Anderson_MRB_1973,Kalmeyer_PRL1987,Balents_Nature2010}.
For classical spins with an antiferromagnetic coupling, 
a geometric frustration suppresses the long-range ordering,
leading to a degenerate ground state.
The degeneracy can be destroyed by quantum fluctuations,
which emerge not only through 
an interplay of strong geometric frustration, low dimensionality, and small spin,
but also through the application of a magnetic field.
Despite intensive research efforts,
the detailed mechanism of the quantum effects, 
e.g., the ground state property\,\cite{YamashitaS_NatP2008,YamashitaM_NatP2009},
has still been highly controversial.

One macroscopic manifestation of the quantum phenomena is the stabilization of the ``up-up-down" spin structure 
under a magnetic field,
predicted for a two-dimensional (2D) triangular-lattice Heisenberg antiferromagnet (TLHAF) with a small spin\,\cite{Nishimori_JPSJ1986,Chubukov_JPCM1991}.
In a magnetization process, the nonclassical anomaly appears as a plateau in a finite field range
at one-third of the saturation magnetization $M_\mathrm{s}$, hereafter referred to as the $M_\mathrm{s}/3$ plateau.
In a classical picture, a monotonic increase in the magnetization is expected up to $M_\mathrm{s}$.
A number of theoretical approaches for explaining the quantum mechanism of the $M_\mathrm{s}$/3 plateau have been proposed\,\cite{Nikumi_JPSJ1993,Alicea_PRL2009,
Takano_JPCS2011,Farnell_JPCM2009,Honecker_JPCM1999,Honecker_JPCM2004,Sakai_PRB2011}.
Thus far, however, few numbers of definite experimental results reserved judgment on the issue.
This is mainly due to the experimental difficulty in growing the model material, 
let alone in observing the $M_\mathrm{s}$/3 plateau purely driven by quantum fluctuations.
In fact, most of the TLHAFs ever studied, such as Cs$_2$CuBr$_4$,\cite{Ono_PRB2003,Fortune_PRB2009} have a distorted triangular lattice,
which induces an antisymmetric Dzyaloshinsky-Moriya (DM) interaction.

It is believed that the spin state in the lower-field range above the higher edge field of the $M_\mathrm{s}$/3 plateau is the $2\,{:}\,1$ canted coplanar state that is a continuous variant of the up-up-down state\,\cite{Chubukov_JPCM1991,Nikumi_JPSJ1993,Alicea_PRL2009,
Takano_JPCS2011,Farnell_JPCM2009}.
However, whether the $2\,{:}\,1$ canted coplanar state is stable up to the saturation or a new quantum spin state emerges in higher-field range is still unclear\,\cite{Nikuni_JPSJ1995}.
To conduct the quantitatively verification of the theory and the elucidation of the high-field spin state, the experimental realization of  2D $S\,{=}\,1/2$ Heisenberg antiferromagnet on a uniform triangular lattice is necessary.

Recently, we have demonstrated that Ba$_3$CoSb$_2$O$_9$ enables us to
experimentally investigate a 2D $S\,{=}\,1/2$ TLHAF system\,\cite{Shirata_PRL2012}.
Magnetic Co$^{2+}$ ions form uniform triangular-lattice layers parallel to the $ab$-plane\,\cite{Treiber_ZAAC1982,Doi_JPCM2004}.
Since the magnetic layers are separated by nonmagnetic layers consisting of the Sb$_2$O$_9$ double octahedron and Ba$^{2+}$ ions, 
the intralayer exchange interaction is expected to be dominant.
On the other hand, the antiferromagnetic ordering at the N\'eel temperature $T_\mathrm{N}$\,$\approx$\,3.8\,K\,\cite{Doi_JPCM2004,Shirata_PRL2012} is due to the weak interlayer exchange interaction.
The DM interaction is absent in Ba$_3$CoSb$_2$O$_9$
because of the highly symmetric crystal structure with the space group $P6_3/mmc$.
The effective magnetic moment of Co$^{2+}$ ions, which possess the true spin $S\,{=}\,3/2$, 
can be described by the pseudospin-1/2 at low temperatures well below
$|\lambda|$/$k_\mathrm{B}$\,$\approx$\,250\,K ($\lambda$: spin orbit coupling constant)\,\cite{Lines_PR1963,Shiba_JPSJ2003}.
Thus, the $M_\mathrm{s}$/3 plateau observed using Ba$_3$CoSb$_2$O$_9$ powder samples in Ref.\,\onlinecite{Shirata_PRL2012}
could be attributed only to the quantum fluctuations.

Many experimental spin-1/2 systems are composed of Cu$^{2+}$ ion that has true spin $S\,{=}\,1/2$. However, it is impossible to realize a regular triangular lattice composed of Cu$^{2+}$ ions, because the orbital degeneracy cannot be lifted in trigonal crystal field, and thus, the trigonal (or hexagonal) crystal lattice is unstable at low temperatures. Therefore, Ba$_3$CoSb$_2$O$_9$ is a rare experimental system of $S\,{=}\,1/2$ antiferromagnet on a uniform triangular lattice, and thus, it is important to perform more precise experiments on Ba$_3$CoSb$_2$O$_9$.

In this letter, we present single-crystal studies of Ba$_3$CoSb$_2$O$_9$,
performed by the high-field magnetization and electron spin resonance (ESR) measurements.
For magnetization processes,
a quantum $M_\mathrm{s}/3$ plateau is clearly observed for $H$\,$\parallel$\,$ab$,
whereas for $H$\,$\parallel$\,$c$, the magnetization curve exhibits a cusp near $M_\mathrm{s}/3$.
We have found a small magnetization anomaly near $(3/5)M_\mathrm{s}$ for $H$\,$\parallel$\,$ab$, which indicates the emergence of a new high-field phase stabilized by the quantum fluctuation.
ESR has proven to be an excellent probe of collective excitation modes with a high sensitivity,
which could provide a significant clue to the configuration of spin triangles. 
As shown below, the ESR modes observed for $H$\,$\parallel$\,$c$,
together with the magnetization process depending on the field direction, 
indicate that Ba$_3$CoSb$_2$O$_9$ is close to the ideal 2D $S\,{=}\,1/2$ TLHAF with a weak easy-plane anisotropy.


Single crystals of Ba$_3$CoSb$_2$O$_9$ were grown from the melt using a Pt tube as a crucible. 
Ba$_3$CoSb$_2$O$_9$ powder, which was prepared by the same procedure described in the previous paper\,\cite{Shirata_PRL2012}, 
was packed into the Pt tube. The temperature of the furnace was reduced from 1700 to 1600\,$^{\circ}$C for three days.
High-field magnetization processes at pulsed magnetic fields up to 47\,T were measured at $T$\,=\,1.3\,K ($<$\,$T_\mathrm{N}$\,$\approx$\,3.8\,K)
at the Institute for Solid State Physics, University of Tokyo.
High-frequency ESR measurements at pulsed magnetic fields up to 13\,T with fixed frequencies ranging from 55 to 405\,GHz
were performed in the temperature range of $1.6-20$\,K,
at the Institute for Materials Research, Tohoku University. 
Usually, the ESR signal of Co$^{2+}$ in the octahedral environment is hard to observe at high temperatures because of the short spin-lattice relaxation time. Therefore, we performed the ESR measurements at helium temperatures, where the spin-lattice relaxation time becomes long enough to observe well-defined ESR signals, as shown below.
For both measurements, we stacked several pieces of plate-shaped samples 
with typical dimensions of $\sim$\,2\,$\times$\,2\,$\times$\,0.5\,mm$^3$. The wide faces were found to be the $ab$-plane by X-ray single-crystal diffraction.
Magnetic fields were applied parallel to the $ab$-plane and $c$-axis.



\begin{figure}
\begin{center}
\includegraphics[width=0.95\linewidth]{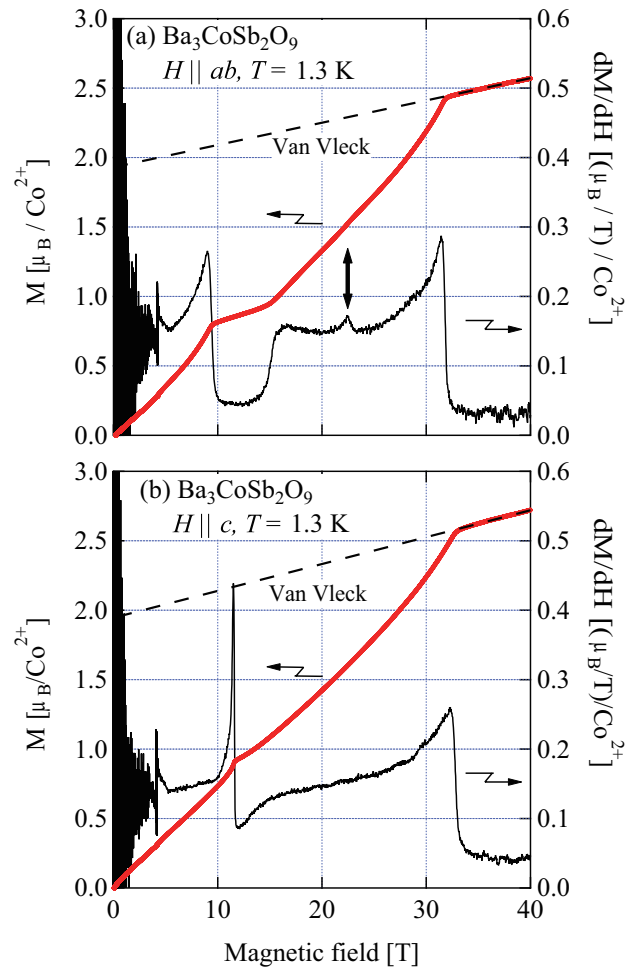}
\end{center}
\vspace{-6mm}
\caption{(Color online) 
Field dependence of the raw magnetization $M_\mathrm{raw}$ (left) and its field derivative $dM_\mathrm{raw}/dH$ (right)
in Ba$_3$CoSb$_2$O$_9$ single crystals, 
measured at 1.3\,K for (a) $H$\,$\parallel$\,$ab$ and (b) $H$\,$\parallel$\,$c$. 
Dashed lines denote the Van Vleck paramagnetism.
A double-headed arrow in (a) indicates an anomaly near 3/5\,$M_\mathrm{s}$, 
possibly related to the spin rearrangement.
}
\vspace{-4mm} 
\label{fig1}
\end{figure}

Figure~\ref{fig1} shows the raw magnetization $M_\mathrm{raw}$ (left)
and its field derivative $dM_\mathrm{raw}/dH$ (right) as a function of $H$ in Ba$_3$CoSb$_2$O$_9$ single crystals. 
The anomaly observed at around 4 T in $dM_\mathrm{raw}/dH$ is due to an instrumental problem.  
The saturation field $H_\mathrm{s}$ is obtained to be 31.9 and 32.8\,T 
for $H$\,$\parallel$\,$ab$ and $H$\,$\parallel$\,$c$, respectively.
The absolute values of the magnetization were calibrated 
with $g$-factors determined as $g\,{=}\,3.84$ and 3.87 for $H$\,$\parallel$\,$ab$ and $H$\,$\parallel$\,$c$, respectively, 
by the present electron paramagnetic resonance (EPR)
measurements performed at 20\,K (see the inset in Fig.~\ref{fig3}).
The magnetic moment of Co$^{2+}$ in Ba$_3$CoSb$_2$O$_9$ is almost isotropic, unlike in typical Co compounds\,\cite{Lines_PR1963}.
The saturation magnetization $M_\mathrm{s}$ is determined as 1.93 and 1.94\,$\mu_\mathrm{B}$/Co$^{2+}$
for $H$\,$\parallel$\,$ab$ and $H$\,$\parallel$\,$c$, respectively.
The $H_\mathrm{s}$ and $M_\mathrm{s}$ values are both consistent with those obtained from our previous results of powder samples\,\cite{Shirata_PRL2012}.
The continuous increase in $M_\mathrm{raw}$ above $H_\mathrm{s}$ is attributed to the large Van Vleck paramagnetism characteristic of Co$^{2+}$ in the octahedral environment\,\cite{Lines_PR1963,Shiba_JPSJ2003}.
From the magnetization slopes above $H_{\rm s}$, the Van Vleck paramagnetic susceptibility is determined as 
$\chi_\mathrm{VV}\,{=}\,1.59\,{\times}\,10^{-2}$ and $1.90\,{\times}\,10^{-2}$\,($\mu_\mathrm{B}$/T)/Co$^{2+}$
for $H$\,$\parallel$\,$ab$ and $H$\,$\parallel$\,$c$, respectively.
It is noted that the magnetization anomalies around $H_\mathrm{s}$ and 
the $M_\mathrm{s}$/3 plateau for $H$\,$\parallel$\,$ab$ are more distinctly observed
as compared with the previous results obtained for powder samples\,\cite{Shirata_PRL2012},
as can also be observed from the sharp anomalies in $dM_\mathrm{raw}/dH$ data.

\begin{figure}
\begin{center}
\includegraphics[width=0.95\linewidth]{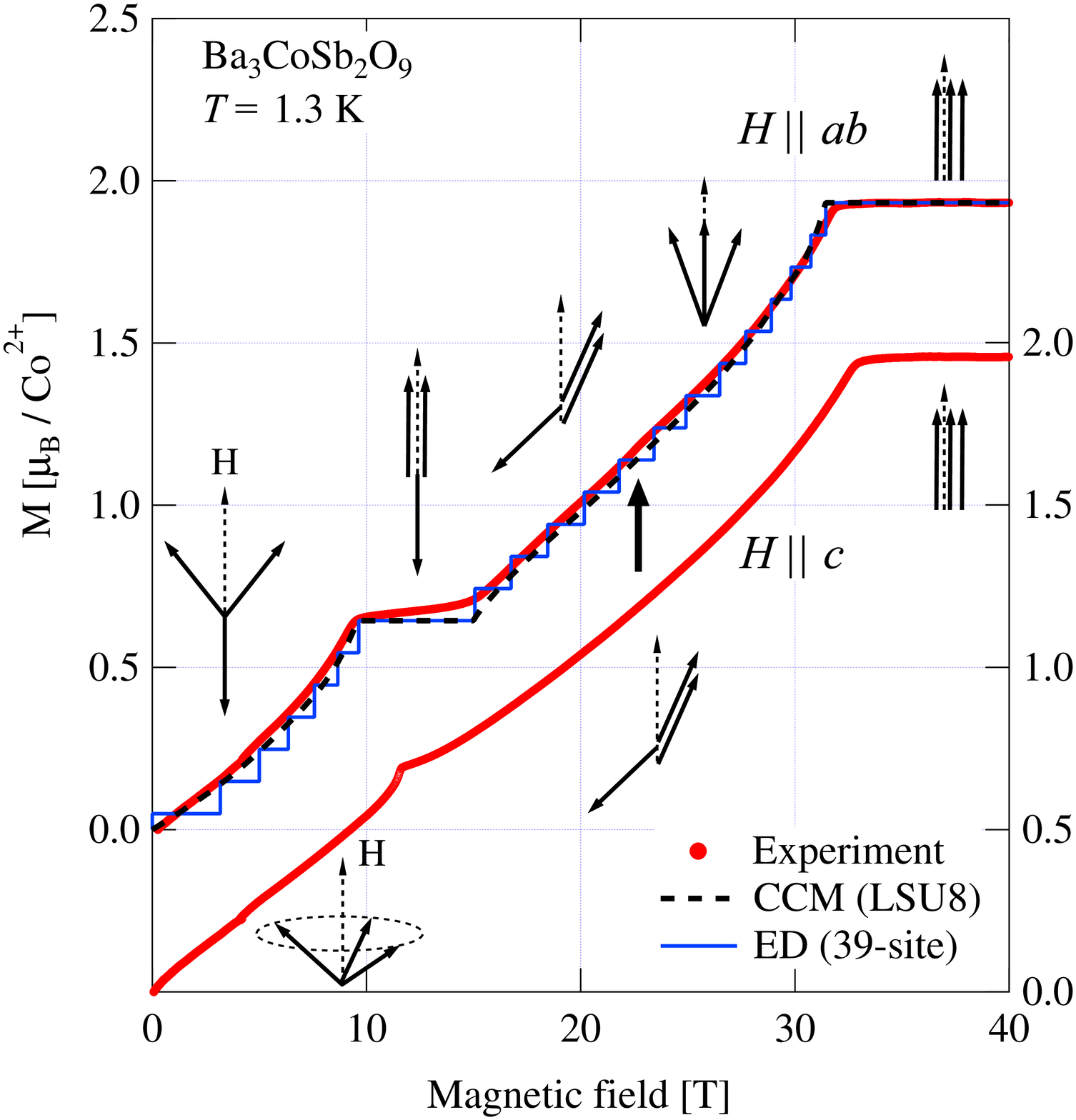}
\end{center}
\vspace{-6mm}
\caption{(Color online) Magnetization curves in Ba$_3$CoSb$_2$O$_9$ at 1.3\,K 
for $H$\,$\parallel$\,$ab$ (left) and $H$\,$\parallel$\,$c$ (right),
corrected for the Van Vleck paramagnetism.
The dashed curve and steplike line denote the results calculated 
by a higher-order coupled cluster method (CCM)\,\cite{Farnell_JPCM2009} 
and exact diagonalization (ED) for a 39-site rhombic cluster\,\cite{Sakai_PRB2011}, respectively.
Spin structures at various magnetic fields are illustrated.} 
\vspace{-4mm}
\label{fig2}
\end{figure}

Figure~\ref{fig2} shows the magnetization curves corrected for the Van Vleck terms.
For $H$\,$\parallel$\,$ab$, a quantum magnetization plateau is clearly observed at $M$\,$\simeq$\,$M_\mathrm{s}$/3.
For comparison, theoretical magnetization curves are also displayed.
Thick dashed and solid lines are the magnetization curves calculated by a higher-order coupled cluster method (CCM)\,\cite{Farnell_JPCM2009} 
and exact diagonalization (ED) for a 39-site rhombic cluster\,\cite{Sakai_PRB2011}, respectively.
Both calculations excellently reproduce the experimental magnetization processes for $H$\,$\parallel$\,$ab$
despite the fact that the adjustable parameters are only $H_\mathrm{s}$ and $M_\mathrm{s}$.
The field range of the experimental $M_\mathrm{s}$/3 plateau is evaluated
to be 0.30\,$\le$\,$H/H_\mathrm{s}$\,$\le$0.47,
which is almost identical to 0.306\,$\le$\,$H/H_\mathrm{s}$\,$\le$0.479 
obtained by CCM\,\cite{Farnell_JPCM2009} and ED\,\cite{Sakai_PRB2011}.

For a 2D $S$\,=\,1/2 TLHAF, the quantum up-up-down spin state is predicted to emerge,
irrespective of the applied field direction.
In Ba$_3$CoSb$_2$O$_9$, however, the magnetization curve for $H$\,$\parallel$\,$c$ exhibits a cusp at $(H, M)\,{=}\,(11.6\,[{\rm T}], 0.66\,[\mu_\mathrm{B}/{\rm Co}^{2+}])\,{\approx}\,(H_\mathrm{s}/3, M_\mathrm{s}/3)$.
The absence of the $M_\mathrm{s}$/3 plateau phase for $H$\,$\parallel$\,$c$
should be ascribed to the weak planar anisotropy in Ba$_3$CoSb$_2$O$_9$.
It has been reported that hexagonal CsCuCl$_3$, which is described as a ferromagnetically stacked triangular-lattice antiferromagnet 
with a small planar anisotropy due to the Dzyaloshinsky-Moriya interaction and the anisotropic exchange interaction\,\cite{Adachi_JPSJ1980}, 
shows similar magnetization behaviors, namely, an ill-defined $M_\mathrm{s}$/3 plateau-like anomaly for $H$\,$\parallel$\,$ab$ and a small jump for $H$\,$\parallel$\,$c$ at around $H_\mathrm{s}$/3\,\cite{Nojiri_CsCuCl3}. 
By analogy with CsCuCl$_3$, the magnetic anisotropy in Ba$_3$CoSb$_2$O$_9$ should be of the easy-plane type,
which will also be confirmed by the analysis of the collective ESR modes, as shown below.
\begin{figure}
\vspace{-2.5mm}
\begin{center}
\includegraphics[width=0.95\linewidth]{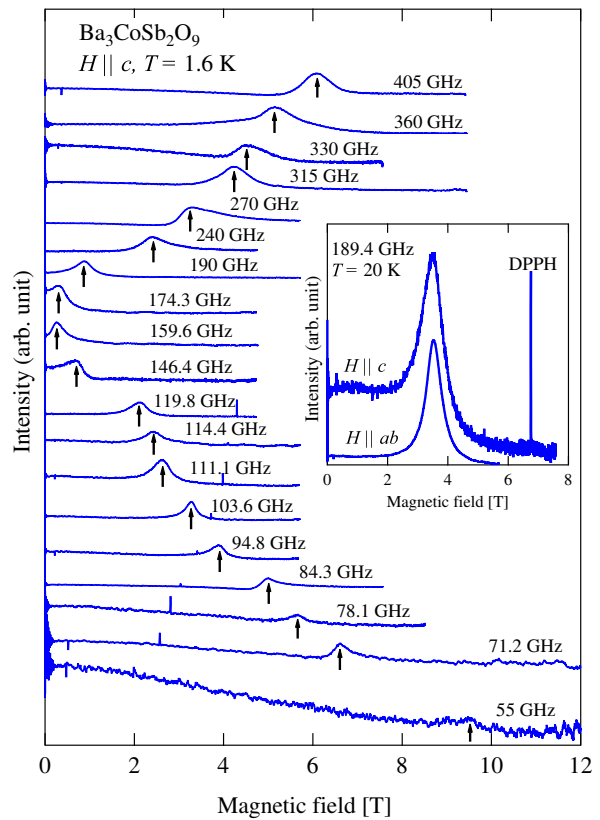}
\end{center}
\vspace{-6mm}
\caption{(Color online) 
ESR spectra of Ba$_3$CoSb$_2$O$_9$ measured at 1.6 K ($<$\,$T_\mathrm{N}$) for $H$\,$\parallel$\,$c$.
Arrows indicate resonance fields.
The inset shows an expanded view of the paramagnetic resonance (EPR) spectra measured 
for $H$\,$\parallel$\,$ab$ and $H$\,$\parallel$\,$c$ at 20 K ($>$\,$T_\mathrm{N}$) and 189.4\,GHz, 
where a sharp line labeled DPPH indicates $g$\,=\,2.} 
\vspace{-4mm}
\label{fig3}
\end{figure}

\begin{figure}
\begin{center}
\includegraphics[width=0.95\linewidth]{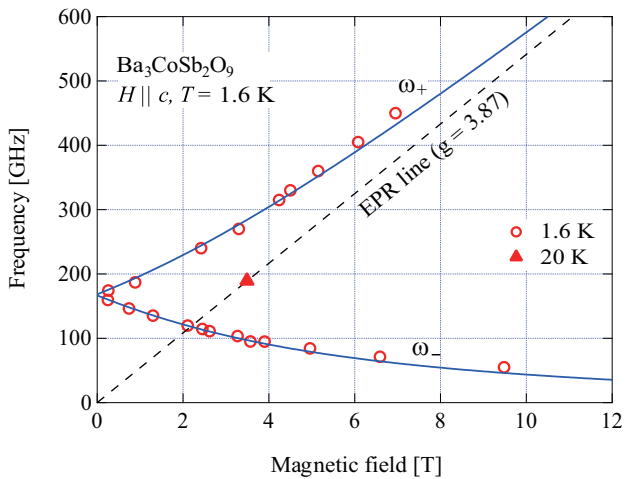}
\end{center}
\vspace{-4mm}
\caption{(Color online) 
Frequency-field diagram of the collective ESR modes in Ba$_3$CoSb$_2$O$_9$ for $H$\,$\parallel$\,$c$.
Open symbols denote the resonance points obtained at 1.6\,K,
where the upper and lower frequency modes are labeled as $\mathrm{\omega_+}$ and $\mathrm{\omega_-}$, respectively.
Solid curves are fits with equation (\ref{equation}) based on a six-sublattice model (see text).
Solid symbols indicate the EPR data measured for $H$\,$\parallel$\,$c$ at 20\,K.
The dashed line is the EPR line described with $g\,{=}\,3.87$.
} 
\vspace{-4mm}
\label{fig4}
\end{figure}

The spin structure expected in each magnetic phase in Ba$_3$CoSb$_2$O$_9$ is illustrated in Fig.~\ref{fig2}.
At zero magnetic field, the spins induce the 120$^{\circ}$ coplanar ordering in the $ab$-plane.
As a magnetic field is applied for $H$\,$\parallel$\,$ab$, 
the quantum up-up-down state with $M\,{=}\,M_\mathrm{s}/3$ is stabilized for 0.297\,$\le$\,$H/H_\mathrm{s}$\,$\le$0.472,
followed by the $2\,{:}\,1$ canted coplanar phase at higher magnetic fields, skipping the up-up-down state. For $H$\,$\parallel$\,$c$, on the other hand, a phase transition occurs at $H\,{\simeq}\,H_\mathrm{s}/3$ from the low-field umbrella structure, which gains the anisotropy energy of the easy-plane type, to the high field $2\,{:}\,1$ canted coplanar one. In the 3D case, this transition accompanies a small magnetization jump\,\cite{Nikumi_JPSJ1993,Nojiri_CsCuCl3}. However, in Ba$_3$CoSb$_2$O$_9$, the magnetization exhibits a cusp anomaly. We infer that the 2D quantum fluctuation changes the magnetization jump into the cusp.

As indicated by a double-headed arrow in Fig.~\ref{fig1}(a), 
one can see a small magnetization anomaly at around 1.5\,[$\mu_\mathrm{B}$/Co$^{2+}$]\,$\approx$\,(3/5)\,$M_\mathrm{s}$ for $H$\,$\parallel$\,$ab$,
which appears as a clear peak in the $dM/dH$ curve.
Such a magnetization anomaly was not detected for $H$\,$\parallel$\,$c$\,\cite{3/5anomaly}.
The magnetization anomaly near $\,(3/5)\,M_\mathrm{s}$ is indicative of the emergence of a new high-field quantum phase. This means that different from the theoretical expectation, the $2\,{:}\,1$ canted coplanar state is not stable up to the saturation. One possibility of the high-field phase is a high-symmetric coplanar phase as shown in Fig.~\ref{fig2}, in which one of the three sublattice spins is aligned along the applied field\,\cite{Nikuni_JPSJ1995}. To determine the high-field spin structure, additional measurements, such as a nuclear magnetic resonance measurement, are required.


Next, we present the results of multifrequency high-field ESR measurements on Ba$_3$CoSb$_2$O$_9$
to provide more clues to the spin configurations and magnetic parameters through the collective magnetic excitations in the ordered state.  
The main panel of Fig.~\ref{fig3} shows the field dependence of ESR spectra at 1.6\,K measured for $H$\,$\parallel$\,$c$ at various fixed frequencies up to 405\,GHz. 
As indicated by arrows, each ESR spectrum in the investigated frequency range has one clear absorption signal,
although the spectrum at 55\,GHz is noisy.
The inset in Fig.~\ref{fig3} shows the paramagnetic resonance (EPR) spectra measured 
for $H$\,$\parallel$\,$ab$ and $H$\,$\parallel$\,$c$ at 20 K ($>$\,$T_\mathrm{N}$) and 189.4\,GHz. 
From the resonance fields of the EPR, the $g$-factors were determined as 
$g\,{=}\,3.84$ and 3.87 for $H$\,$\parallel$\,$ab$ and $H$\,$\parallel$\,$c$, respectively. Such almost isotropic $g$ factor is rare for Co$^{2+}$ in the octahedral environment.

The resonance data obtained are summarized in the frequency-field diagram shown in Fig.~\ref{fig4}.
Open and solid symbols represent the collective ESR modes (1.6\,K) and EPR (20\,K) data, respectively. 
The collective ESR modes consist of two distinct branches labeled as $\mathrm{\omega_+}$ and $\mathrm{\omega_-}$.
At zero magnetic field, the two modes appear to be degenerate 
with an energy gap of approximately 170\,GHz.
As the magnetic field increases, the energy of the $\mathrm{\omega_+}$ mode increases and tends to approach the EPR line with $g$\,=\,3.87,
whereas the energy of the $\mathrm{\omega_-}$ mode decreases monotonically toward zero.
The field evolutions of the $\mathrm{\omega_+}$ and $\mathrm{\omega_-}$ modes for $H$\,$\parallel$\,$c$ are 
characteristic of the triangular-lattice antiferromagnet with the easy-plane anisotropy rather than the easy-axis one\,\cite{Tanaka_JPSJ1988,Palme_SSC1990,Tanaka_JPSJ1992}.
If the anisotropy is of the easy-axis type, no degeneracy of the $\mathrm{\omega_+}$ and $\mathrm{\omega_-}$ modes 
occurs at zero magnetic field.

Here, we analyze the collective ESR modes for $H$\,$\parallel$\,$c$ in Ba$_3$CoSb$_2$O$_9$. 
We consider six sublattices, because the neighboring triangular layers are antiferromagnetically coupled\,\cite{Doi_JPCM2004}. 
The resonance conditions can be derived analytically by solving the torque equations for the six-sublattice model as
\begin{widetext}
\begin{equation}
{\hbar}{\omega}_{\pm}=\sqrt{\left(4J^{\prime}+\frac{9}{2}J\right)\left\{\frac{3\Delta J}{4}+\frac{(8J^{\prime}+9J-6\Delta J)}{2(4J^{\prime}+9J+3\Delta J)^2}\,(g\mu_{\rm B}H)^2\right\}}\ {\pm}\ \frac{9J}{8J^{\prime}+18J+6\Delta J}\,g\mu_{\rm B}H,
\label{equation}
\end{equation}
\end{widetext}
\noindent
where $J$ and $J^{\prime}$ are the intralayer and interlayer exchange constants, respectively, 
and ${\Delta}J$ is the coefficient of anisotropic exchange interaction in the layer defined as ${\Delta}J(S_i^xS_j^x\,{+}\,S_i^yS_j^y)$. The resonance conditions of eq.\,(\ref{equation}) are the same as those obtained from the linear spin wave theory.
Because the intralayer exchange interaction is dominant, the condition $J\,{\gg}\,{\Delta}J, J^{\prime}$ is satisfied in Ba$_3$CoSb$_2$O$_9$.

The intralayer exchange constant $J$ was evaluated to be $J/k_{\rm B}\,{=}\,18.5$\,K 
from the saturation fields $H_{\rm s}$ for $H\,{\parallel}\,ab$ and $H\,{\parallel}\,c$ using the relation $g{\mu}_{\rm B}H_{\rm s}\,{=}\,9J/2$. 
In the analysis, we fixed the value of $J$ and the $g$-factor of 3.87 obtained via the EPR measurement for $H$\,$\parallel$\,$c$.
Solid curves in Fig.~\ref{fig4} are fits obtained using equation (\ref{equation}) 
with $J^{\prime}/k_\mathrm{B}\,{=}\,0.48$\,K and ${\Delta}J/k_\mathrm{B}\,{=}\,1.02$\,K, 
which are sufficiently smaller than $J$. The agreement between the experimental and theoretical results is excellent.
The ESR results indicate that, in Ba$_3$CoSb$_2$O$_9$, the intralayer exchange interaction is dominant over the interlayer one,
and that the magnetic anisotropy is small and is of the easy-plane type, which is in disagreement with the recent report by Zhou {\it et al.}\,\cite{Zhou_PRL2012}.
In Ref.\,\onlinecite{Shirata_PRL2012}, we have proposed that Ba$_3$CoSb$_2$O$_9$ has the easy-axis-anisotropy, 
deduced from the three-step specific heat anomaly around $T_\mathrm{N}$\,$\approx$\,3.8\,K at zero magnetic field. 
However, three specific heat peaks measured by the relaxation method around $T_\mathrm{N}$ were extremely high 
considering that the phase transition is of the second order. 
We infer that the specific heat anomaly arises from a weak first order phase transition accompanied by the latent heat, 
as predicted for the XY and Heisenberg triangular-lattice antiferromagnets\,\cite{Zumbach_PRL1993,Loison_EPJB1998}.   

To conclude, the high-field magnetization and ESR of Ba$_3$CoSb$_2$O$_9$ single crystals have been investigated.
A quantum $M_\mathrm{s}/3$ plateau is clearly observed for $H\,{\parallel}\,ab$,
whereas for $H\,{\parallel}\,c$, the magnetization curve exhibits a cusp near $H_\mathrm{s}/3$. 
The suppression of the $M_\mathrm{s}/3$ plateau for $H\,{\parallel}\,c$ is ascribed to the easy-plane anisotropy.
We have found an small magnetization step at nearly (3/5)\,$M_\mathrm{s}$ for $H\,{\parallel}\,ab$, indicative of a new high-field quantum phase, which has not been predicted so far. 
The expected spin configurations are proposed as shown in Fig.~\ref{fig2}.
Nearly isotropic $g$-factors of 3.84 and 3.87 are obtained for $H$\,$\parallel$\,$ab$ and $H$\,$\parallel$\,$c$, respectively, from the EPR spectrum at 20\,K. 
From the detailed analysis of two distinct ESR modes observed for $H$\,$\parallel$\,$c$
 ($T$\,$<$\,$T_\mathrm{N}$), 
combined with the obtained magnetization processes,
we determined interaction parameters in Ba$_3$CoSb$_2$O$_9$ with accuracy as $J/k_{\rm B}\,{=}\,18.5$\,K, $J^{\prime}/k_\mathrm{B}\,{=}\,0.48$\,K and ${\Delta}J/k_\mathrm{B}\,{=}\,1.02$\,K. 
This result indicates that Ba$_3$CoSb$_2$O$_9$ closely approximates the ideal two-dimensional $S\,{=}\,1/2$ TLHAF. These interaction parameters are necessary for the quantitative discussion of quantum aspects of magnetic excitations such as negative quantum renormalization of excitation energies and the unusual singularity of dispersion relation predicted for the $S\,{=}\,1/2$ TLHAF~\cite{Starykh,Zheng,Chernyshev}.

We would like to thank D. J. J. Farnell, R. Zinke, J. Schulenburg, J. Richter, H. Nakano, T. Sakai and D. Yamamoto for showing us their theoretical calculations of the magnetization processes and for fruitful discussion.
This work was supported by a Grant-in-Aid for Scientific Research (A) 
from the Japan Society for the Promotion of Science, and the Global COE Program 
``Nanoscience and Quantum Physics'' at Tokyo Tech. 
funded by the Ministry of Education, Culture, Sports, Science and Technology of Japan. 
H. T. was supported by a grant from the Mitsubishi Foundation.

\end{document}